\documentclass[prd,%preprint,
twocolumn,superscriptaddress,preprintnumbers,nofootinbib,10pt,aps
]{revtex4-2}

\usepackage{amsmath,amssymb,bm,graphicx,siunitx,xcolor}
\definecolor{rossoCP3}{cmyk}{0,.88,.77,.40}
\newcommand{\CLASS}{\texttt{CLASS}}
\newcommand{\lcdm}{\ensuremath{{\Lambda\text{CDM}}}}

\begin{document}
\title{\color{rossoCP3} Massive neutrinos and interacting dark matter look alike through the lens
  of lensing}

\author{\bf Luis A. Anchordoqui}

\affiliation{Department of Physics \& Astronomy,  Lehman College, City University of
  New York, NY 10468, USA
}

\affiliation{Department of Physics, 
 Graduate Center,  City University of
  New York,  NY 10016, USA
}

\affiliation{Department of Astrophysics,
 American Museum of Natural History, NY
 10024, USA
}

\author{\bf Danny Marfatia}

\affiliation{Department of Physics \& Astronomy, University of Hawaii at Manoa, 2505 Correa Rd., Honolulu,
HI 96822, USA}

\author{\bf Jorge F. Soriano}
\affiliation{Department of Physics \& Astronomy,  Lehman College, City University of
  New York, NY 10468, USA
}

\begin{abstract}
 \noindent We demonstrate that the suppression in the lensing power spectrum of the cosmic microwave background (CMB) caused by massive neutrinos can be mimicked by dark matter-baryon interactions at the precision of next-generation CMB experiments. 
 Thus, a determination of neutrino masses from the CMB lensing power spectrum may be compromised.
 We illustrate the degeneracy for a dark matter-proton cross section $\propto v^{-4}$, which arises in the $t$-channel exchange of an ultralight mediator in the nonrelativistic limit. 
\end{abstract}  

\maketitle

\section{Introduction}

The distribution of matter in the Universe carries abundant
information about primordial density perturbations and the forces
that have shaped cosmological evolution. Mapping this distribution
is one of the main goals of modern cosmology. Gravitational lensing
of the cosmic microwave background (CMB) probes the matter between Earth
and the surface of last scattering by leveraging the 
exquisitely well-known statistical properties of the relic photons. As CMB photons travel to Earth from the last
scattering surface, they are deflected by the intervening matter, which
distorts the observed pattern of CMB anisotropies and modifies their
statistical properties. Gravitational lensing shifts
the apparent arrival direction of CMB photons and breaks the
primordial statistical isotropy of the unlensed CMB; lensing thus correlates previously independent Fourier modes of the CMB temperature power spectrum. These correlations can be used to make a map of the gravitational potential that altered the photon paths. Since the gravitational potential encodes information
about the formation of structure in the Universe, CMB
lensing encodes plenty of statistical information about
large-scale structure. In fact, the structure along
the line-of-sight is sensitive to various cosmological
parameters.

It is well known that the CMB and matter power spectra constrain
the sum of the neutrino masses $\sum m_\nu$ because while neutrinos freestream, they suppress power on scales smaller than the horizon when neutrinos become nonrelativistic~\cite{Eisenstein:1997jh}.
It is also well known that these power spectra provide
strong constraints on dark matter (DM) interactions with baryons~\cite{Chen:2002yh,Dvorkin:2013cea,  Gluscevic:2017ywp,Xu:2018efh,Slatyer:2018aqg,Buen-Abad:2021mvc}. These interactions lead to
momentum and heat exchange between the ordinary matter and cold dark matter (CDM) components. The momentum exchange rate between the baryon and DM
fluids acts as friction on the latter, damping the growth of
perturbations in the energy density of the DM fluid, thereby
suppressing structure formation. All in all, a characteristic
signature of DM-baryon (DMb) interactions is the suppression of power
on small scales. 

It is of interest to investigate the
degeneracy between the suppression of power caused by neutrino mass
and that caused by the DMb interaction. One can be played against the
other, weakening the sensitivity of CMB experiments to neutrino
masses.

Before proceeding, we pause to note that neutrino mass and the dynamics of dark energy are strongly correlated.
If the dark energy equation of state as a function of redshift satisfies $w(z) \geq -1$ for all $z$, then neutrino mass limits become more restrictive~\cite{Vagnozzi:2018jhn,Jiang:2024viw}. However, DESI data hint towards dark energy with $w (z) < -1$ for $z \agt 0.5 $~\cite{DESI:2025zgx}. For a detailed account of a physically motivated scenario, see~\cite{Bedroya:2025fwh}. Then, upper limits on $\sum m_\nu$ are significantly relaxed. In our investigation we consider a constant dark energy density given by a cosmological constant $\Lambda$. 

In this paper, we study how the unknown DMb scattering
cross section could impact the determination of the neutrino mass scale
from the lensing power spectrum. The layout is as follows. 
In Section~\ref{sec:1} we recount the effects of neutrino mass on the power spectra.
In Section~\ref{sec:2} we summarize our scheme for the DMb interaction and
review existing upper limits on the DMb scattering cross section. In
Section~\ref{sec:3} we use numerical methods to characterize the parameter space in which the DMb
interaction can mimic the effects of $\sum m_\nu$ on
structure growth. We conclude in Section~\ref{sec:4}.

\section{Neutrino masses}
\label{sec:1}

%One such parameter is the sum of the neutrino masses $\sum m_\nu$. 
%For the purpose of the discussion thatfollows, it is instructive to briefly summarize how a measurement of $\sum m_\nu$ could be carried out. 
If the Universe has a standard thermal
history, with neutrino decoupling before $e^+/e^-$ annihilation, we
can infer the present
temperature of the relic neutrino background with remarkable accuracy:
\begin{equation}
T_\nu^0 = (4/11)^{1/3} \ T_\gamma^0 = (1.676 \pm 0.001) \times
  10^{-4}~{\rm eV}  \,,
\label{Tnu}
\end{equation}
where $T^0_{\gamma}$ is the temperature of the CMB photons~\cite{ParticleDataGroup:2024cfk}. It is common
to express the total contribution of neutrinos to the energy density
as
\begin{equation}
  \Omega_\nu h^2 \sim \frac{\sum m_\nu}{93.14~{\rm eV}}\,,
\label{Omeganu}
\end{equation}
where $\Omega_i$ is the fraction of the critical density contributed by the $i^{\rm th}$ matter species ($\nu$ = neutrinos, $\gamma$ = photons,
$m$ = nonrelativistic matter) and $h$ is the Hubble constant in units of 100~km/s/Mpc~\cite{Lesgourgues:2006nd}. In Eq.~(\ref{Omeganu}) we neglected the following contribution of neutrinos lighter than $T_\nu^0$ to the energy density:
\begin{equation}
\Omega_{\nu, {\rm rel}} h^2 =  (4/11)^{4/3} \ \Omega_\gamma h^2= (0.6414 \pm
0.001) \times 10^{-5} \,,
\end{equation}
per flavor.

We assume that the standard thermal history has a power-law spectrum of
primordial adiabatic density fluctuations $P(k)$. Early
universe fluctuations, under the influence of gravity, 
grow into large-scale structures. However, the large momentum of
cosmological neutrinos prevents them from falling into these gravitational
wells on small scales, smearing out any small-scale density variations
and effectively acting as a damping force on structure
formation. This damping effect due to neutrino {\it free-streaming} suppresses the growth of cosmic structures on scales
smaller than the distance neutrinos can travel in the time it takes
for them to become nonrelativistic (nr). 
For each neutrino of mass $m_\nu$, the power spectrum is suppressed for angular wavenumbers
larger than the free-streaming wavenumber~\cite{Lesgourgues:2006nd},
\begin{equation}
  k_{\rm nr} \sim 0.018 \ (m_\nu/{\rm eV})^{1/2} \ \Omega_m^{1/2} h \
{\rm Mpc}^{-1} \, .
\end{equation}  
Cosmological measurements of the small-scale suppression~\cite{Hu:1997mj}, 
\begin{equation}
  \Delta P/P \sim - 8 \ \Omega_\nu/\Omega_m \, ,
\label{DeltaP}
\end{equation}  
provide an indirect method of determining neutrino masses given
Eq.~(\ref{Omeganu})~\cite{Hu:1997mj}. The
linear estimate in Eq.~(\ref{DeltaP}) is a reasonable first-order
approximation for $0 < \Omega_\nu/\Omega_m <
0.07$~\cite{Lesgourgues:2006nd}, which corresponds to $\sum m_\nu \alt
1.5~{\rm eV}$.

The {\it Planck} CMB power spectra suggest slightly more lensing than
predicted by the benchmark $\Lambda$CDM cosmology~\cite{Planck:2018vyg}. Assuming only neutrino
masses suppress power, the Planck Collaboration reported a 95\%~CL
upper limit from CMB angular power spectra, $\sum m_\nu < 0.26~{\rm eV}$. This constraint is improved to $\sum m_\nu < 0.12~{\rm
  eV}$ when data from baryon acoustic oscillations (BAO) are included. Next generation CMB experiments are expected to produce high-fidelity maps over large regions
of the sky, and improve the signal-to-noise of the {\it Planck} lensing maps~\cite{Planck:2018lbu} by more than an order of
magnitude. The angular power spectrum of the lensing potential derived from these observations is expected to resolve
differences in the neutrino mass scale of order $\sigma (\sum m_\nu) =
20~{\rm meV}$~\cite{TopicalConvenersKNAbazajianJECarlstromATLee:2013bxd,Yu:2018tem,Chang:2022tzj}.

\section{DM\lowercase{b} interactions}
\label{sec:2}

We extend $\Lambda$CDM by including DMb
interactions. Rather than focusing on specific DM models, we adopt a phenomenological
description of the momentum transfer cross section,
\begin{equation}
  \sigma = \sigma_{\rm DMb}\ v^n\,,
\end{equation}
where $v$ is the magnitude of the relative velocity between the
incoming DM particle  and the baryon,
$\sigma_{\rm DMb}$ is the velocity-stripped momentum transfer cross
section, and $n$ depends on the type of interaction. %This phenomenological description allows us to simplify the model implementation and enlarges the applicability of the results when mapping to concrete dark matter models.
There are a number of well-motivated choices for $n$ (between $-4$ and 6), some of which  have been studied in a cosmological setting~\cite{Dvorkin:2013cea,Xu:2018efh}.
For example, the
velocity-independent $n = 0$ cross section characterizes 
contact interactions with low-momentum
transfer~\cite{Chen:2002yh}, and is instantiated by DM with a magnetic dipole moment, 
 while $n = -2 $ occurs if the DM has an
electric dipole moment~\cite{Sigurdson:2004zp}.

We focus on $n=-4$ as it illustrates the interplay between interacting DM and massive neutrinos most clearly. This velocity dependence naturally arises in the $t$-channel exchange of an ultralight scalar or vector mediator in the nonrelativistic limit.
The matrix element is ${\cal{M}}\sim1/(t - m^2)$, where $m$ is the mediator mass and $t \simeq -|\bar{q}|^2 \simeq - (\mu v)^2$ is the momentum transfer in the nonrelativistic limit, with $\mu$ the reduced mass of the DM-baryon system. Thus, for an ultralight mediator, ${\cal{M}} \sim t^{-1}$, and the cross section $\sigma\sim |{\cal{M}}|^2 \sim v^{-4}$. 
The canonical example is that of millicharged DM~\cite{Holdom:1985ag,Dvorkin:2013cea}. However, in this case, the region of parameter space of interest is in tension with observations~\cite{Mahdawi:2018euy}. 

\begin{figure}[t]
\includegraphics[width=\linewidth]{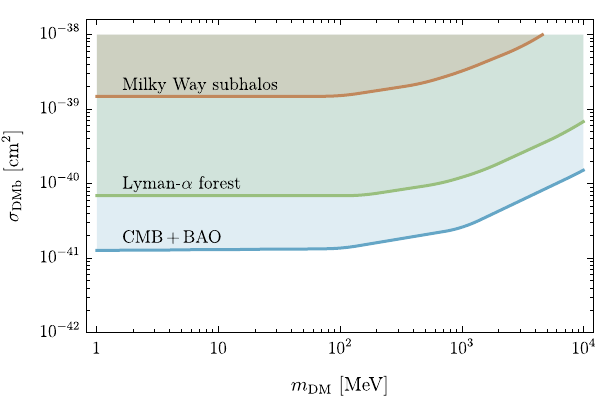}
      \caption{95\%~CL limits on  $\sigma_{\rm DMb}$ for $n=-4$, from CMB+BAO spectra, Milky Way subhalos and the Lyman-$\alpha$ forest~\cite{Buen-Abad:2021mvc}. \label{fig:bounds}}
      \end{figure}

In Fig~\ref{fig:bounds} we provide 
limits derived from CMB+BAO spectra, the abundance of Milky Way
subhalos, and the Lyman-$\alpha$ forest~\cite{Buen-Abad:2021mvc}. The motivation for considering these data is as follows.

{\it (i)}~CMB spectra provide constraints on the strength of the DMb interaction, primarily by detecting the resulting suppression of structure formation~\cite{Chen:2002yh,Dvorkin:2013cea, Gluscevic:2017ywp,Xu:2018efh,Slatyer:2018aqg,Buen-Abad:2021mvc}.

{\it (ii)}~The number and distribution of DM subhalos in the Milky
Way's halo are sensitive to DMb interactions~\cite{Escudero:2018thh,DES:2020fxi}. DM 
scattering off baryons can alter the distribution of DM within the subhalos, can lead to the formation of ``cores'' in the centers of subhalos or deplete their numbers in the inner regions of the Milky Way. 
By comparing predictions from simulations with observed Milky Way
satellite galaxies, which act as tracers of these DM subhalos, upper bounds on the strength of DMb interactions can be derived.

{\it (iii)}~The Lyman-$\alpha$ forest consists of numerous absorption lines in the
spectra of distant quasars due to intervening gas
clouds. If DM interacts with baryons, it can heat or cool the gas in
the intergalactic medium. This alters the gas temperature and thus
affects the width of the Lyman-$\alpha$ absorption lines. By comparing
the observed line widths with theoretical predictions from
hydrodynamical simulations, upper bounds on the strength
of the DMb interaction are obtainable~\cite{
Chen:2002yh,Dvorkin:2013cea,
Gluscevic:2017ywp,Xu:2018efh,Slatyer:2018aqg,Buen-Abad:2021mvc}. A
point worth noting is that Lyman-$\alpha$ constraints are subject to systematic uncertainties in the modeling of the intergalactic medium, such as its thermal history~\cite{Hui:2016ltb}. 

Note that upcoming observations of the 21-cm signal will be able to probe $10^{-43} \alt \sigma_{\rm DMb}/{\rm cm^2} \alt 10^{-44}$  for $m_{\rm {DM}}< 100$~MeV and $n=-4$~\cite{Rahimieh:2025fsb}.

\begin{figure*}[t]\centering
  \hfill\includegraphics[width=0.45\linewidth]{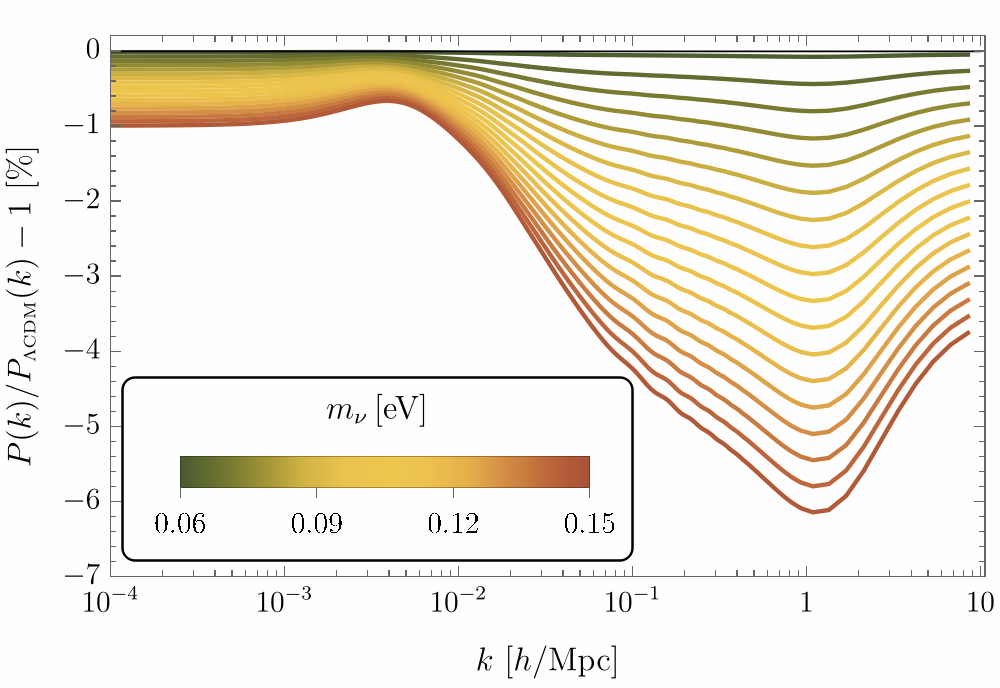}\hfill
  \includegraphics[width=0.45\linewidth]{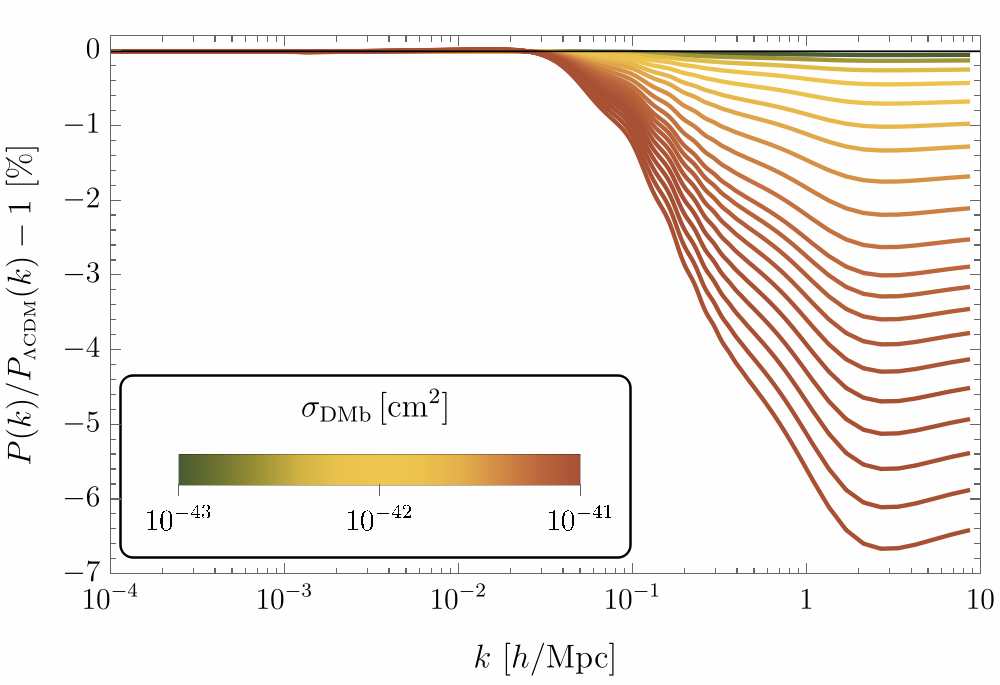}\hfill\,
  \caption{Relative differences in the matter power spectra for
    neutrino masses (left) and DMb interactions with $m_\mathrm{DM}=\qty{0.1}{GeV}$ (right), with respect to $\Lambda$CDM with $\sum m_\nu=0.06\,\mathrm{eV}$ and the parameters in Eq.~(\ref{eq:baselcdm}) set to $\{0.02242,0.11933,0.6766,0.0561,3.047,0.9665\}$~\cite{Planck:2018vyg}. } \label{fig:mpk}
\end{figure*}

\section{$\bm{\sum m_\nu}$ or DM\lowercase{b} scattering?} 
\label{sec:3}

To simulate the gravity-driven evolution of particles over cosmic time we utilize the publicly available Boltzmann solver \CLASS~\cite{Blas:2011rf,Lesgourgues:2011rh}, using the {\tt idm}~\cite{Becker:2020hzj} parameters for DM-proton interactions from version \texttt{3.3.2}.
We have
verified that for protons, the results of \CLASS' \texttt{idm} parameter choices are
consistent with those from the {\tt class}$_-${\tt dmb} package developed
in~\cite{Buen-Abad:2021mvc}, which is based on an older version of \CLASS. To capture
nonlinearities of the gravitational clustering that forms DM halos, filaments, and voids  we use  
{\tt HALOFIT}~\cite{Smith:2002dz,Bird:2011rb}.

The $\Lambda$CDM model can be described by the six base parameters,
\begin{equation}
 \{\omega_b,\omega_c,h,\tau_{\rm reio},\ln (10^{10}A_s),n_s\}\,,\label{eq:baselcdm}\end{equation}
where
$\omega_b \equiv \Omega_b h^2$ is the
baryon density, $\omega_c = \Omega_{\rm CDM} h^2$ is the CDM density,
$\tau_{\rm reio}$
is the Thomson scattering optical depth due to reionization, $A_s$ is
the normalization of the primordial power spectrum of scalar perturbations, and $n_s$ is its spectral
index. Beyond the base parameters, we set the number of effective
relativistic degrees of freedom $N_{\rm eff} = 3.046$. 
The three neutrino species are approximated as two
massless states and a single massive neutrino of mass $m_\nu$, which we take between 0.06~eV and 0.1~eV.

The DMb model is described by two additional parameters, $\sigma_\mathrm{DMb}$ and the DM mass $m_\mathrm{DM}$. We assume that all the DM is interacting.

\begin{figure*}\centering
  \hfill\includegraphics[width=0.45\linewidth]{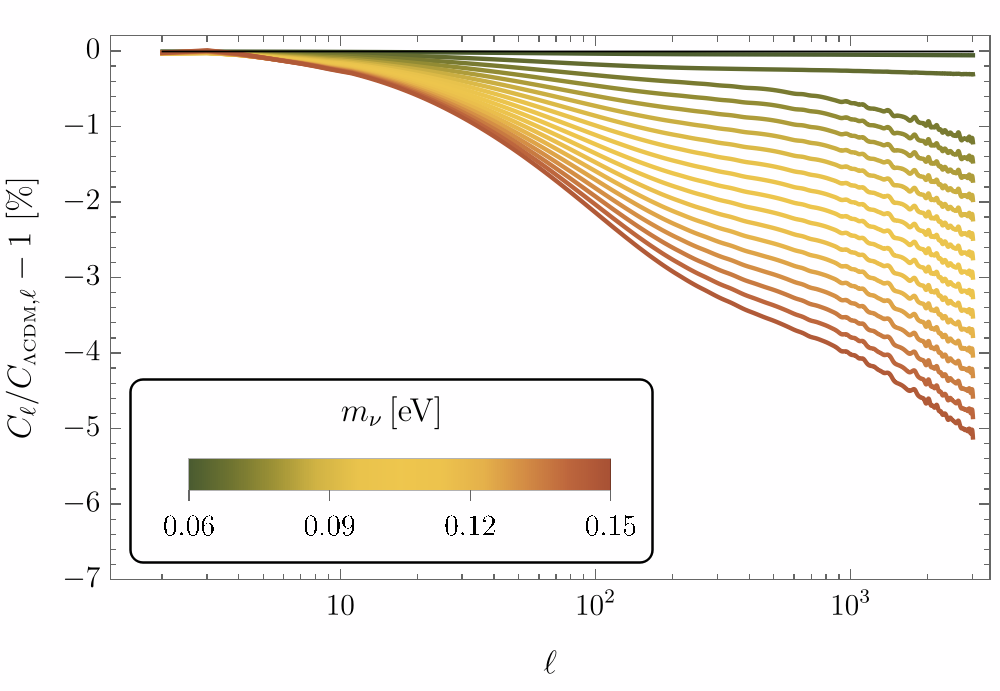}\hfill
  \includegraphics[width=0.45\linewidth]{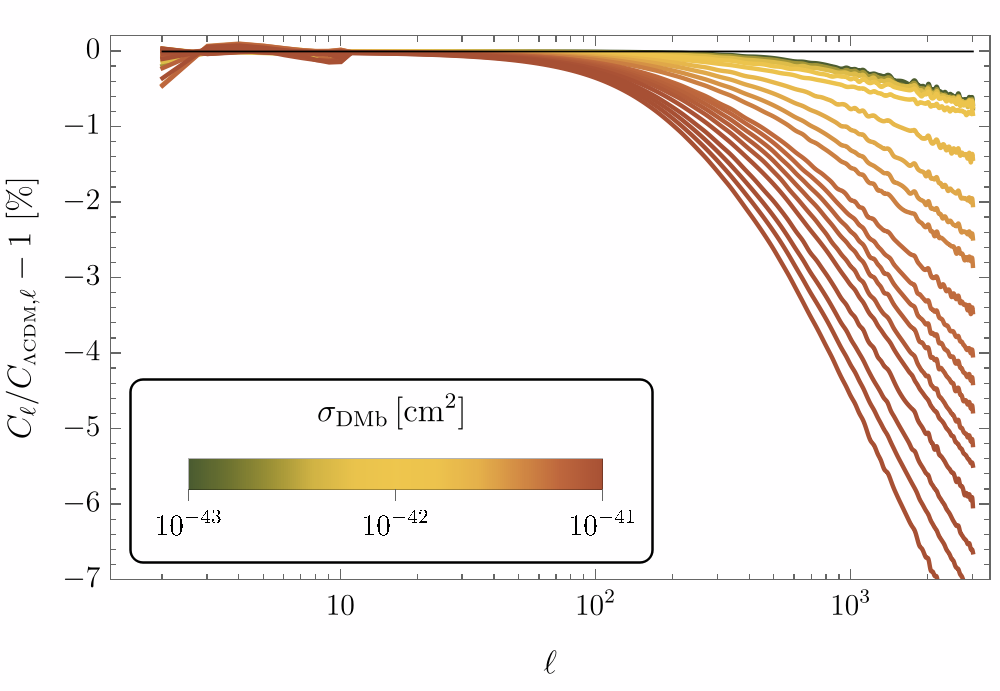}\hfill\,
  \caption{Relative differences in the lensing power spectra for
    neutrino masses (left) and DMb
    interactions (right), with the same parameter choices as in Fig.~\ref{fig:mpk}. } \label{fig:lcl}
\end{figure*}

In Fig.~\ref{fig:mpk} we illustrate the suppression of power caused by increasing either the neutrino mass or the DMb cross section. Although the suppressed spectra are not identical, they exhibit some commonalities that lead to a correlation between $\sum m_\nu$ and $\sigma_\mathrm{DMb}$ when performing a fit to data.
In Fig.~\ref{fig:lcl} we show the impact of these scenarios on the lensing power spectrum.

\begin{table}[t]\centering 
    \begin{tabular}{ccc}\toprule
        likelihood & $\sum m_\nu = 0.06\,\mathrm{eV}$ & Model A \\\hline
        \texttt{Planck\textunderscore highl\textunderscore TTTEEE} & $2353.22$ & $2353.92$ \\
        \texttt{Planck\textunderscore lowl\textunderscore EE     } & $397.46$  & $397.72$ \\
        \texttt{Planck\textunderscore lowl\textunderscore TT     } & $22.11$   & $22.08$ \\
        \texttt{Planck\textunderscore lensing     }                & $9.30$    & $9.56$ \\
        \texttt{bao\textunderscore angular        }                & $33.77$   & $37.23$ \\\hline
        Total $\chi^2$                                                      & $2815.87$ & $2820.53$ \\\toprule
    \end{tabular}
     \caption{Minimum $\chi^2$ for $\Lambda$CDM with $\sum m_\nu=0.06\,\mathrm{eV}$ and $\sum m_\nu=0.1\,\mathrm{eV}$ (model A) from an analysis of Planck and BAO data.}
     \label{tab:lcdm:chis}
\end{table}

To explore whether DMb interactions could spoil the detection of neutrino mass from the lensing power spectrum, we study the similarity between the lensing power spectra with and without DMb interactions for different neutrino masses. In particular, consider two models of the Universe: model A, with massive neutrinos and no DMb interactions, and model B, with massive neutrinos lighter than those in model A, and with DMb interactions. In model B both mechanisms suppress the matter power spectrum, while in model A neutrinos are solely responsible. 

Either model may be the true model of our Universe. Analyses assuming model A when B is the true model (or vice versa) may lead to incorrect conclusions. We now illustrate how the similarity of the power spectra in models A and B may spoil an independent determination of neutrino masses.
To do so, we take model A to have $\sum
m_\nu=0.1\,\mathrm{eV}$, and model B to have $\sum m_\nu=0.06\,\mathrm{eV}$, corresponding to the atmospheric neutrino mass scale.%, and scan the parameter space generated by $\sigma_{\rm DMb}$ and $m_{\rm DM}$.

We analyze these models in two ways. First, we study the extent to which they are compatible with existing data. In particular, we compare the level of agreement of models A and B with the current accepted \lcdm\ model, in which $m_\nu=\qty{0.06}{eV}$. This will determine what combinations of $m_\mathrm{DM}$ and $\sigma_\mathrm{DMb}$ are excluded by current cosmological data. Second, we compare models A and B with each other using their lensing power spectra. This will demonstrate how the effects on the lensing power spectrum of neutrino masses and of interacting dark matter look alike.

To confirm the compatibility of these models with existing data, we perform an analysis using \texttt{MontePython}~\cite{Brinckmann:2018cvx,Audren:2012wb} with the following likelihoods: \texttt{Planck\textunderscore highl\textunderscore TTTEEE}, \texttt{Planck\textunderscore lowl\textunderscore TT}, \texttt{Planck\textunderscore lowl\textunderscore EE} and \texttt{Planck\textunderscore lensing} for the Planck temperature and polarization spectra with lensing, and \texttt{bao\textunderscore angular} for BAO. The best fits for the \lcdm\  model ($\sum m_\nu=0.06\,\mathrm{eV}$), and for model A ($\sum m_\nu=0.1\,\mathrm{eV}$), which share the same six base parameters are provided in Table~\ref{tab:lcdm:chis}. It is well established that smaller neutrino masses produce better fits, and we find that 
$\sum m_\nu=0.1\,\mathrm{eV}$ is compatible with a benchmark $\Lambda$CDM model with $\sum
m_\nu=0.06\,\mathrm{eV}$ at $2.15\sigma$.

\begin{figure*}\centering
\includegraphics[width=0.48\linewidth]{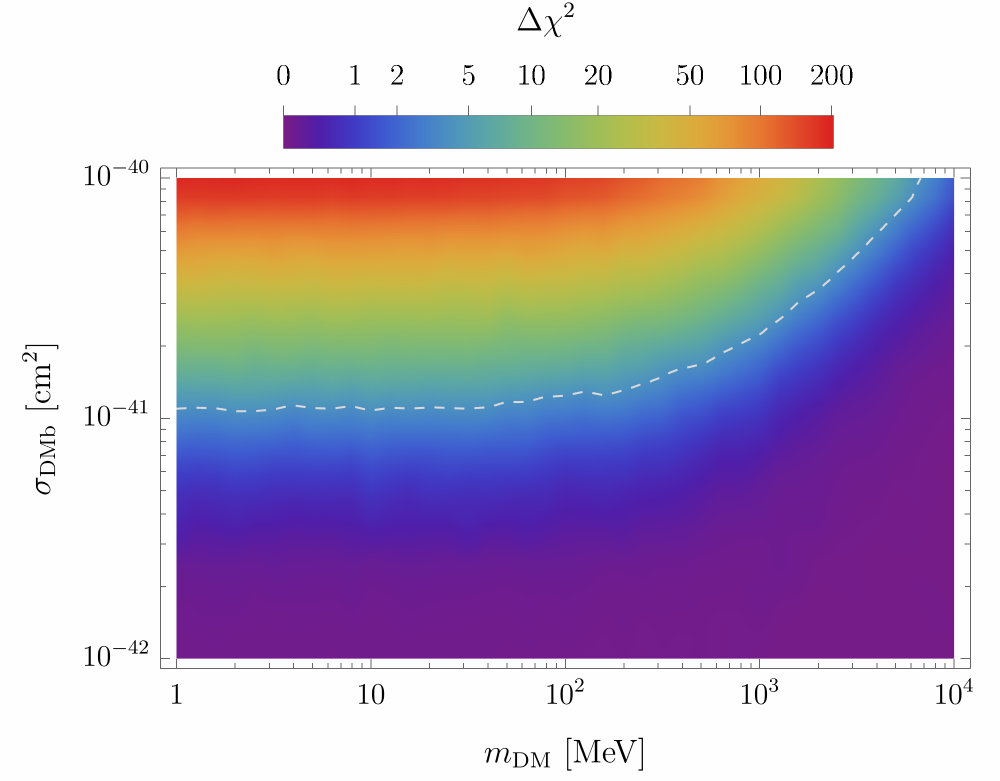}
  \includegraphics[width=0.48\linewidth]{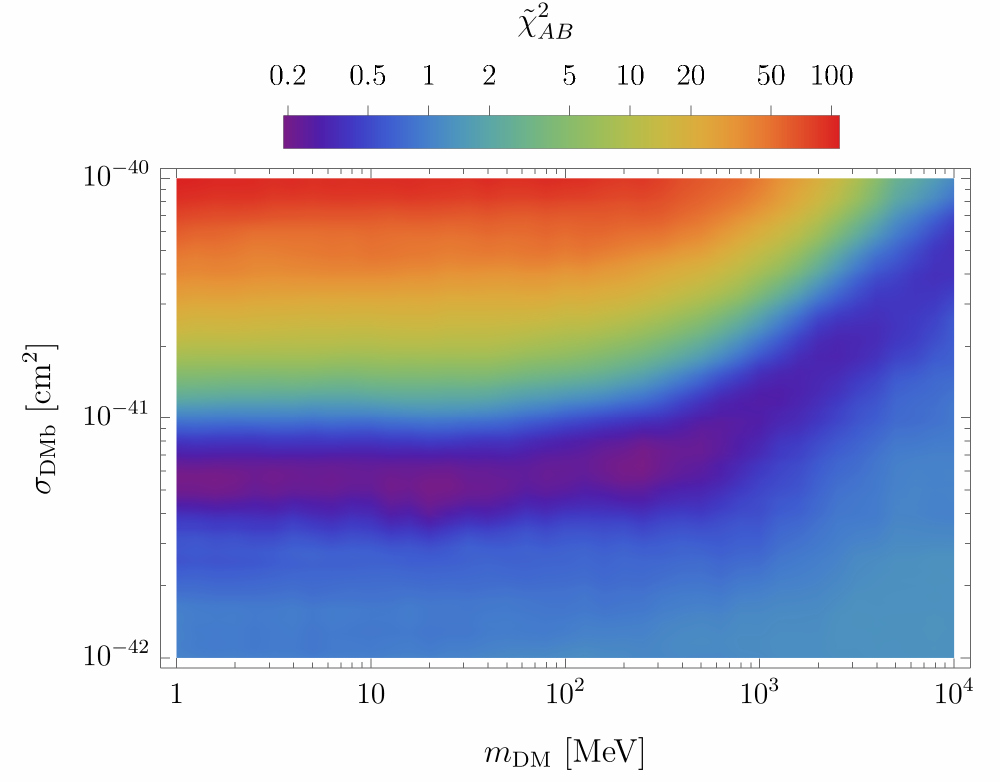}\hfill
  \caption{Values of $\Delta\chi^2$ (left) and $\tilde\chi^2_{AB}$ (right) in the ($m_\mathrm{DM},\sigma_\mathrm{DMb}$) plane. The dashed curve in the left panel corresponds to $\chi^2_A-\chi^2_\lcdm=4.66$, and points below it provide a better fit to Planck and BAO data than model~A. The colors represent the entire range of values in the numerical analysis.  }\label{fig:chi2}
\end{figure*}

\begin{figure}[t]\centering
  \includegraphics[width=\linewidth]{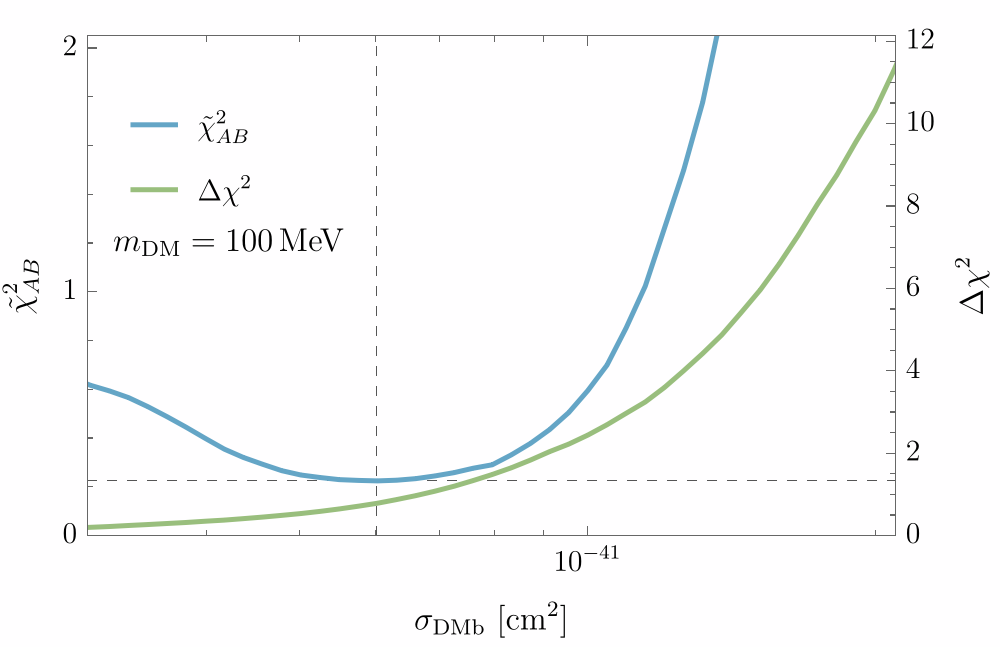}
  \caption{Values of $\tilde\chi^2_{AB}$ (blue, left axis) and $\Delta\chi^2$ (green, right axis) for $m_\mathrm{DM}=\qty{100}{MeV}$ and different values of $\sigma_\mathrm{DMb}$.}\label{fig:chi2-common}
\end{figure}

We now perform a similar analysis for model B for each choice of $(m_\mathrm{DM},\sigma_\mathrm{DMb})$ by minimizing the $\chi^2$ with respect to the six base parameters in Eq.~(\ref{eq:baselcdm}).  As a metric for this comparison, we define the difference $\Delta\chi^2=\chi^2_B-\chi^2_\lcdm$, where $\chi^2_\lcdm=2815.87$ (from Table~\ref{tab:lcdm:chis}).
We show values of $\Delta\chi^2$ in the  ($m_\mathrm{DM},\sigma_\mathrm{DMb}$) plane in the left panel of Fig.~\ref{fig:chi2}, where the dashed line corresponds to $\chi^2_A-\chi^2_\lcdm=4.66$. Note that points below the dashed curve provide a better fit to the data than model A. The green curve in Fig.~\ref{fig:chi2-common} shows  $\Delta\chi^2$ for $m_\mathrm{DM}=\qty{100}{MeV}$. Since $\Delta\chi^2\geq0$ throughout the parameter space, \lcdm\ is a better fit than model~B. $\Delta\chi^2 \to 0$ only as $\sigma_\mathrm{DMb}\to0$.

We now compare models~A and~B with each other using their lensing power spectra. Using \CLASS\ we obtain the lensing power spectra for models~A and~B, $C_\ell^{A}$ and $C_\ell^{B}(\sigma_{\rm DMb},m_{\rm DM})$ respectively, in the interval 
$\ell\in[50,3000]$. We use the weighted average of their squared differences,
\begin{equation}
  \chi_{AB}^2(\sigma_{\rm DMb},m_{\rm DM})=\sum_\ell  {\frac{\left(C_\ell^{A} - C_\ell^{B}(\sigma_{\rm DMb},m_{\rm DM})\right)^2}{\sigma_\ell^2}},\label{chi2ab}
\end{equation}
as a measure of their similarity.
The CMB-S4 projected uncertainties $\sigma_\ell$ are 
extracted from Fig.~49 of Ref.~\cite{CMB-S4:2016ple}. For all
multipoles, the projected uncertainties are larger than the limit imposed by cosmic variance. 

\begin{figure}[t]\centering
\includegraphics[width=\linewidth]{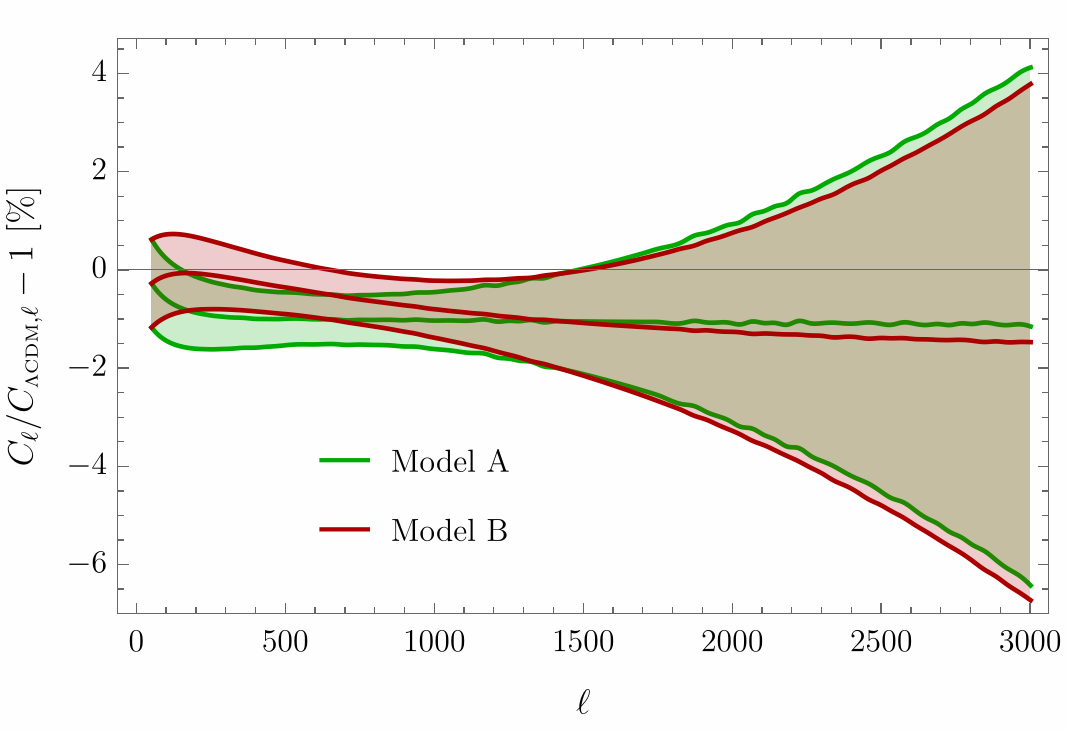}
\caption{Relative differences in the lensing power spectra for model A with $\sum m_\nu=\qty{0.1}{eV}$ and $\sigma_\mathrm{DMb}=0$, and model B with $\sum m_\nu=\qty{0.06}{eV}$, $m_\mathrm{DM}=\qty{100}{MeV}$ and $\sigma_\mathrm{DMb}\approx\qty{6e-42}{cm^2}$. We use symmetric uncertainty bands at $C_\ell\pm\sigma_\ell$. }
\label{fig:lcl-unc}
\end{figure}

In the right panel of Fig.~\ref{fig:chi2} we show the values of $\tilde\chi_{AB}^2=\chi_{AB}^2/N$ in the 
($m_\mathrm{DM},\sigma_\mathrm{DMb}$) plane, where $N=2951$ is the number of terms in the sum over $\ell$. This makes $\tilde\chi_{AB}^2$ analogous to a reduced chi-squared, so $\tilde\chi_{AB}^2\lesssim1$ means that models A and B are indistinguishable within uncertainties. In a sizable region (purple and dark blue) of parameter space, the differences between models A and B are minor. 
The blue curve in Fig.~\ref{fig:chi2-common} shows 
$\tilde\chi_{AB}^2$ for $m_\mathrm{DM}=\qty{100}{MeV}$.
 It has a minimum at $\sigma_\mathrm{DMb}\approx\qty{6e-42}{cm^2}$. We choose this value to illustrate in Fig.~\ref{fig:lcl-unc}, the similarities between the lensing power spectra of models A and B, within the uncertainty bands derived from $\sigma_\ell$.

\begin{figure}[t]\centering
  \includegraphics[width=\linewidth]{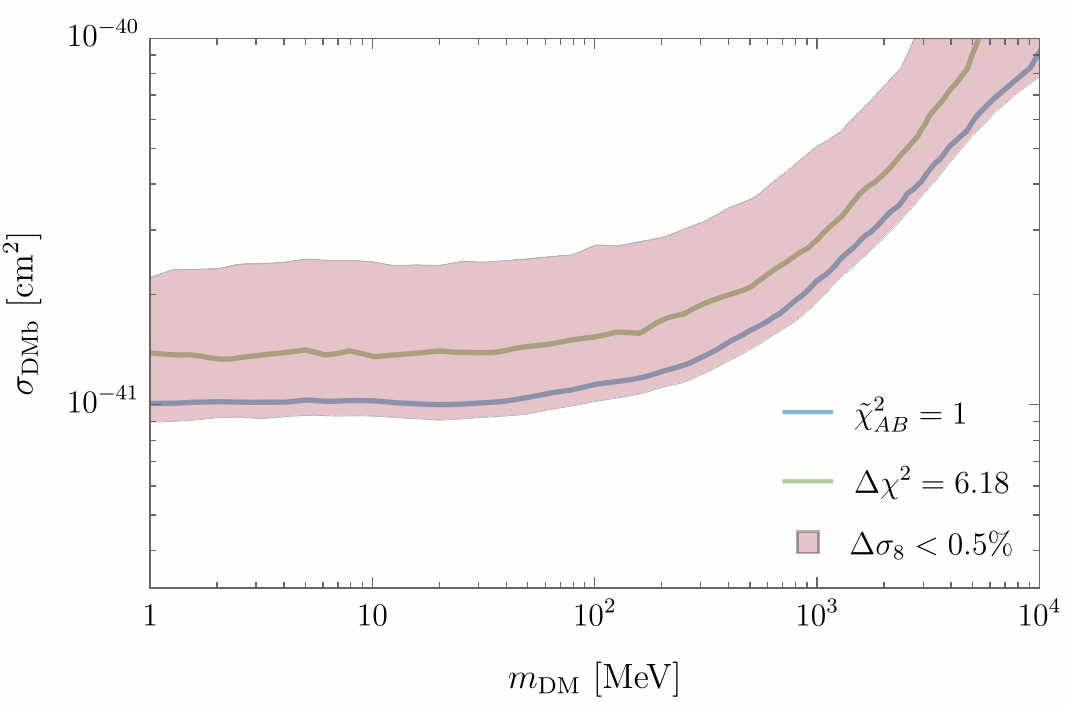}
  \caption{Upper bounds on $\sigma_\mathrm{DMb}$ that satisfy $\Delta\chi^2<6.18$ (green) and
  $\tilde\chi^2_{AB}<1$ (blue). In the shaded band, the values of $\sigma_8$ for the two models differ by less than $0.5\%$.}\label{fig:chi2-common-2d}
\end{figure}

From Figs.~\ref{fig:chi2} and~\ref{fig:chi2-common}, it is evident that a region exists in which \emph{(i)} model B is not excluded by cosmological data, and \emph{(ii)} models A and B are indistinguishable. We further illustrate this in Fig.~\ref{fig:chi2-common-2d} by showing the upper bounds on $\sigma_\mathrm{DMb}$ in the $(m_\mathrm{DM},\sigma_\mathrm{DMb})$ plane that satisfy $\tilde\chi^2_{AB}<1$ and $\Delta\chi^2<6.18$ (corresponding to model B being compatible with current data within $2\sigma$).
We find that if $\sigma_\ell$ is reduced by a factor of two, 
then $\tilde\chi^2_{AB} \gtrsim 1$ in the entire range of $m_\mathrm{DM}$, thus enabling discrimination between models A and B.

Large-scale structure data can achieve high precision determinations of $\sigma_8$ (the root-mean-square amplitude of linear matter fluctuations in spheres of radius 8$h^{-1}\rm{Mpc}$ at $z=0$), which is a sensitive probe of our models. In the shaded region of Fig.~\ref{fig:chi2-common-2d}, the values of $\sigma_8$ for the two models differ by less than $0.5\%$. 
Current $68\%\,\mathrm{C.\,L.}$ uncertainties on measurements of $\sigma_8$ are 
$\sim 2\%$~\cite{Wright:2025xka}, and the uncertainties expected with lensing experiments may reach $\sim 0.3\%$~\cite{CMB-S4:2016ple}.

%We can similarly estimate the increase of precision required for models A and B to become incompatible within the methods discussed in this work. A decrease in the value of $\sigma_\ell$ to $\alpha\,\sigma_\ell$ (with $\alpha<1$) increases $\tilde\chi^2_{AB}$ by $\alpha^{-2}$, reducing the size of the $\tilde\chi^2_{AB}\leq1$ region. This creates two tensions. First, decreasing $\alpha$ reduces the upper limit in $\sigma_\mathrm{DMb}$ for which $\tilde\chi^2_{AB}<1$ (blue curve in Fig.~\ref{fig:chi2-common-2d}). Eventually, this constraint becomes incompatible with the $\Delta\sigma_8<0.5\%$ condition. This occurs around $\alpha=0.8$. Second, for a further decrease in $\alpha$, the region $\tilde\chi^2_{AB}<1$ disappears completely. This occurs at $\alpha\approx0.5$. This gives an estimate of the precision in lensing measurements required to disentangle the similarities between models A and B.

We stress that the possibility of interacting dark matter in the parameter space region with $\tilde\chi_{AB}^2 < 1$ would spoil claims about high-precision lensing measurements of neutrino mass.

\section{Conclusions}
\label{sec:4}

The suppression in the CMB lensing power spectrum due to neutrino mass can be mimicked by dark matter-proton interactions with a cross section $\propto v^{-4}$. This compromises the measurement of neutrino masses from the lensing power spectrum. To end on an optimistic note, because DMb interactions can only enhance the power suppression, any upper bound on neutrino masses that neglects DMb interactions is conservative.

\newpage
\section*{Acknowledgements}

The work of L.A.A. is supported by the U.S. National Science
Foundation (NSF Grant PHY-2412679), and he thanks the Harvard Swampland Initiative for hospitality.
The work  of D.M. is supported by the U.S. Department of Energy under Grant No.~DE-SC0010504.

\end{document}